\documentstyle[aps,prd,eqsecnum,12pt]{revtex}

\begin{document}
\draft
\tighten
\title{Extended Kelvin theorem in relativistic magnetohydrodynamics }

\author{ Jacob D. Bekenstein and Asaf Oron}
\address{\it The Racah Institute of Physics, Hebrew University of
Jerusalem,\\ Givat Ram, Jerusalem 91904, Israel
\\E-mails:
bekenste@vms.huji.ac.il and asafo@alf.fiz.huji.ac.il}

\maketitle
\bigskip\bigskip\bigskip
\begin{abstract}
We prove the existence of a generalization of Kelvin's circulation theorem
in general relativity which is applicable to perfect isentropic
magnetohydrodynamic flow. The argument is based on a new version of the
Lagrangian for perfect magnetohydrodynamics.  We illustrate the new conserved
circulation with the example of a relativistic magnetohydrodynamic flow
possessing three symmetries.
\end{abstract}
\bigskip
\section{Introduction}

It is hard to overstate the important role that Kelvin's theorem on the
conservation of circulation of a simple perfect fluid has played in the
development of hydrodynamics.  Among other things it provided the basis for
the discussion of potential flows, and showed that isolated
vortices should exist, and that they obey the Helmholtz laws, etc.  On
the other hand, Kelvin's theorem is fragile: as soon as dissipation comes
in, it breaks down.  And when the body force per unit mass of fluid is not a
gradient, as happens for the Lorentz force in magnetohydrodynamics (MHD),
Kelvin's theorem ceases to apply. 

Often fluids in the real world carry magnetic fields.  For example, the fluid
at the Earth's core, the plasma in the sun and pulsars, the ionized gas in
interstellar space and in supernova remnants, the plasma in intergalactic
space, and many others, all carry magnetic fields.  These are important
systems for which we need insights of the sort Kelvin's theorem bestowed on
ordinary fluid dynamics.  Can we extend Kelvin's theorem to MHD  ? 

The first such extension was found two decades ago by E.
Oron$^{\!\!\!\cite{bekenstein1}}$ working with the formalism of relativistic
perfect MHD.  This is a circulation theorem for a hybrid velocity-magnetic
field combination.  Oron's derivation assumes both stationary symmetry and
axisymmetry, while it is well known that Kelvin's theorem requires neither of
these.  Yet it has not proved possible to rid Oron's result of the symmetry
assumptions.

In the present paper we follow, on the wake of our earlier
paper,$^{\!\!\!\cite{bekenstein2}}$ a different route.  We use the least
action principle to give a rather straightforward  existence proof for a
generically conserved hybrid velocity--magnetic field circulation in general
relativistic MHD which does not depend on the presence of spacetime
symmetries.   Recently Els\"asser$^{\!\!\!\cite{els}}$ has given a related
result which he obtains by direct manipulation of the relativistic MHD
equations.

A formal introduction to relativistic MHD is given by
Lichnerowicz.$^{\!\!\!\cite{lichnerowicz}}$  As mentioned, we approach the
whole problem not from equations of motion, but from the least action
principle.  In special relativity Penfield$^{\!\!\!\cite{penfield}}$ proposed
a perfect fluid Lagrangian which admits vortical isentropic unmagnetized
flow.  The early general relativistic Lagrangian of Taub$^{\!\!\!\cite{taub}}$
as well as the more recent one by Kodama et. al$^{\!\!\!\cite{kodama}}$
describe only irrotational perfect fluid flows. The Lin
device$^{\!\!\!\cite{lin}}$ to include vortical flows is incorporated by
Schutz$^{\!\!\!\cite{schutz}}$ in his Lagrangian.
Carter$^{\!\!\!\cite{carter}}$ has introduced a relativistic Lagrangian for
particle-like motions from which the properties of fluid flows, including
vortical ones, can be inferred.  However, it does not correspond in detail to
the MHD paradigm.  Achterberg$^{\!\!\!\cite{achterberg}}$ proposed a general
relativistic MHD action, which, however, describes only ``irrotational''
flows.  Thompson$^{\!\!\!\cite{thompson}}$ used this Lagrangian in the
extreme relativistic limit. Heyl and Hernquist$^{\!\!\!\cite{heyl}}$ 
modified it to include QED effects. In this paper we follow
Schutz's$^{\!\!\!\cite{schutz}}$ approach while supplementing it by the
introduction of magnetic fields

In Sec.~II.A we review the relativistic MHD equations.  In Sec.~II.B we
describe our Lagrangian,$^{\!\!\!\cite{bekenstein2}}$ and show that it gives
the correct Maxwell and fluid equations, while in Sec.~II.C  we recover the
relativistic Euler equation from it. In Sec.~III.A we derive the general form
of the conserved circulation, while in Sec.III.B  we illustrate it with the
special case of a MHD flow endowed with three symmetries. 

\bigskip
\section{Relativistic Action Principle}

In this section we construct a Lagrangian density for MHD flow in general
relativity (GR). Greek indices run from 0 to 3.  The coordinates are denoted
$x^\alpha=\left(x^0,x^1,x^2,x^3\right)$; $x^0$ stands for time.  A comma
denotes  the usual partial derivative; a semicolon covariant
differentiation.   Our signature is $\{-,+,+,+\}$. We take $c=1$.

\subsection{Relativistic MHD Equations}
Our first step is enumerating all the correct general relativistic (GR)
equations for MHD. These were  developed
by Lichnerowicz,$^{\!\!\!\cite{lichnerowicz}}$ Novikov  and  
Thorne,$^{\!\!\!\cite{novikov}}$
Carter,$^{\!\!\!\cite{carter}}$ Bekenstein  and  
Oron$^{\!\!\!\cite{bekenstein1}}$ and
others. 

The first equation states the conservation of  the number of particles (we do
not consider particle annihilation or creation processes),
\begin{equation}
\label{gr4}
N^\alpha{}_{;\alpha }=\left( nu^{\alpha }\right) _{;\alpha }=0,
\end{equation}
where $N^\alpha$ is the particle number 4--current density, $n$ the particle
proper number density and $u^\alpha$ the fluid 4--velocity field normalized by
$u^\alpha u_\alpha=-1$.  We consider flows which are inviscid and adiabatic,
and therefore  $s$,  the entropy per particle,  is conserved along flow lines:
\begin{equation}
\label{gr_entropy}
\left( sN^{\alpha }\right) _{;\alpha }=0\qquad {\rm or} \qquad u^\alpha
s_{,\alpha}=0.
\end{equation}

The energy momentum tensor for the magnetized fluid is obtained by
adding the electromagnetic energy--momentum tensor to that of an ideal
fluid:
\begin{equation}
\label{gr5}
T^{\alpha \beta }=pg^{\alpha \beta }+\left( p+\rho \right) u^{\alpha }
u^{\beta }+  (4\pi)^{-1}(F^{\alpha\gamma} F^\beta{}_\gamma - {\scriptstyle
1\over\scriptstyle 4} F^{\gamma\delta} F_{\gamma\delta}\,
g^{\alpha\beta}).
\end{equation}
Here $\rho$ denotes the fluid's  proper energy density (including rest
mass) and $p$ the scalar pressure (assumed isotropic), while
$F^{\alpha\beta}$ denotes the electromagnetic field tensor which satisfies
the  GR Maxwell's equations
\begin{eqnarray}
\label{Maxwell1}
F^{\alpha\beta}{}_{;\beta} &=& 4\pi j^\alpha
\\
 F_{\alpha\beta,\gamma} +
F_{\beta\gamma,\alpha}+F_{\gamma\alpha,\beta}  &=& 0,
\label{Maxwell2}
\end{eqnarray}
where $j^\alpha$ denotes the electric 4--current density.

The magnetic Euler equation for the fluid is derived from the vanishing  
covariant divergence law $T^{\alpha \beta }{}_{;\beta}=0$:
\begin{equation}
\label{em_conservation}
(\rho+p)u^\beta u^\alpha{}_{;\beta} = - (g^{\alpha\beta}+u^\alpha
u^\beta)p_{,\beta} +(4\pi)^{-1} F^{\alpha\beta}
F_{\beta}{}^\gamma{}_{;\gamma}.
\end{equation}
The term $a^\alpha\equiv u^\beta u^\alpha{}_{;\beta}$ stands for the fluid's
acceleration 4--vector. The use of covariant derivatives and curved
metric ensures that effects of gravitation are automatically included. In view
of Eq.~(\ref{Maxwell1})  the above yield the MHD Euler equation
\begin{equation}
\label{grmag_euler}
(\rho+p)a^\alpha = - h^{\alpha\beta} p_{,\beta} +
F^{\alpha\beta} j_\beta,
\end{equation}
where $h^{\alpha\beta} $ is the  projection tensor
\begin{equation}
\label{h}
 h^{\alpha\beta}\equiv g^{\alpha\beta} + u^\alpha u^\beta.
\end{equation}

The Euler equation describes a general  electromagnetic field carrying
flow. One needs an additional condition to distinguish MHD flow from all 
others.  For any flow carrying an electromagnetic field,
the (antisymmetric) Faraday tensor $F_{\alpha \beta }$ may be split into
electric and magnetic vectors with respect to the flow:
\begin{eqnarray}
\label{gr1}
E_{\alpha }&=&F_{\alpha \beta }u^{\beta }
\\
\label{gr2}
B_{\alpha }&=&{}^*F_{\beta\alpha}u^\beta\equiv {\scriptstyle
1\over\scriptstyle 2}\epsilon_{\beta\alpha
\gamma \delta }\,F^{\gamma \delta } u^{\beta }.
\end{eqnarray}
Here $ \epsilon _{\alpha \beta \gamma \delta } $ is the
Levi-Civita totally antisymmetric tensor ($\epsilon_{0123}=(-g)^{1/2}$ with
$g$ denoting the determinant of the metric $g_{\alpha\beta}$) and
${}^*F_{\alpha\beta}$ is the dual of $F_{\alpha\beta}$.  In a frame comoving
with the fluid, these 4--vectors have only spatial parts which correspond to
the usual ${\bf E}$ and ${\bf B}$, respectively.  One can use
Eqs.~(\ref{gr1}-{\ref{gr2}) to express $F_{\alpha\beta}$ using those 4--vectors
\begin{equation}
F_{\alpha\beta}=u_\alpha E_\beta - u_\beta E_\alpha
+\epsilon_{\alpha\beta\gamma\delta}u^\gamma B^\delta
\label{inversion}
\end{equation}
For an infinitely conducting (perfect
MHD) fluid, the electric field in the fluid's frame must
vanish,$^{\!\!\!\cite{lichnerowicz}}$ i.e.,
\begin{equation}
\label{gr3}
E_{\alpha }=F_{\alpha \beta } u^\beta = 0.
\label{freeze}
\end{equation}
This corresponds to the usual MHD condition  ${\bf E} + {\bf v} \times {\bf
B}=0$.

\subsection{Relativistic Lagrangian density and equations of motion }

We now propose a Lagrangian density for GR MHD flow based on
Schutz's$^{\!\!\!\cite{schutz}}$ Lagrangian density for pure fluids in GR:
\begin{equation}
{\cal L}=-\rho(n,s) -(16\pi)^{-1}F_{\alpha \beta }F^{\alpha \beta }+
\phi N^{\alpha }{}_{;\alpha }+\eta \left( sN^{\alpha }\right) _{;\alpha
}+\lambda \left( \gamma N^{\alpha }\right) _{;\alpha }+\tau^{\alpha
}F_{\alpha\beta }N^{\beta }.
\label{Lagrangian}
\end{equation}
As shown below, our Lagrangian reproduces Eqs.~(\ref{gr4}-\ref{gr_entropy}),
(\ref{Maxwell1}-\ref{grmag_euler}) and (\ref{gr3}).  Here $\phi$ is the
Lagrange multiplier associated with the conservation of particle number 
Eq.~(\ref{gr4}) viewed as a constraint,   $\eta $ is that multiplier
associated with the adiabatic flow constraint, Eq.~(\ref{gr_entropy}), and
$\lambda $ is  that associated with the conservation along the flow of
Lin's$^{\!\!\!\cite{lin}}$ quantity $\gamma$.  $\tau_{\alpha}$ is a quartet of
Lagrange multipliers which enforce the field freezing condition
Eq.~(\ref{gr3}).  We  should interject that $\tau^\alpha$ is not determined
uniquely, as we discuss in Sec.~III.

We view $\gamma$, $N^\alpha$ and $s$ as the independent fluid variables, while
$n$ and $u^\alpha$ are determined  by the obvious relations
\begin{equation}
\label{grem0}
-N^{\alpha }N_{\alpha }=n^2; \qquad u^\alpha = n^{-1}\,N^\alpha.
\end{equation}
Some authors prefer to include in the Lagrangian the constraint
$N^{\alpha }N_{\alpha }+n^2=0$, which stands for the normalization of
the fluid's 4--velocity. However we choose to impose this constraint later
and thus remain with a simpler Lagrangian.

We can now vary the Lagrangian with respect to the independent variables.  
Variation of $\phi$ recovers the conservation of
particles $N^\alpha{}_{;\alpha}=0$.  Variation of
$\lambda$ with subsequent use of the previous result yields
\begin{equation}
\gamma_{,\alpha }u^{\alpha }=0.
\label{gr_lambda}
\end{equation}
If we vary Lin's $ \gamma  $ we get
\begin{equation}
\lambda_{,\alpha }u^{\alpha }=0.
\label{gr_gamma}
\end{equation}
These  results inform us that $\gamma$ and $\lambda$ are both locally 
conserved along the flow.  In view of the thermodynamic relation
$n^{-1}(\partial \rho/\partial s)_n = T$, with $T$ the locally
measured fluid temperature, variation  of
$s$ gives
\begin{equation}
u^{\alpha }\eta _{,\alpha }=-T.
\label{gr_s}
\end{equation}

We now vary $N^\alpha$ using the obvious consequence of Eq.~(\ref{grem0}),
\begin{equation}
\label{grem2}
\delta n=-u_{\alpha }\delta N^{\alpha },
\end{equation}
together with the thermodynamic relation$^{\!\!\!\cite{novikov}}$ involving
the specific enthalpy $\mu$,
\begin{equation}
\label{gr_mu1}
\mu \equiv (\partial\rho/\partial n)_s =n^{-1}\,(\rho +p).
\end{equation}
We thus get the most important equation herein:
\begin{equation}
\mu u_{\alpha }=\phi _{,\alpha }+s\eta _{,\alpha }+\gamma
\lambda _{,\alpha }+
\tau^{\beta }F_{\alpha\beta }.
\label{gr_varu}
\end{equation}
By contracting Eq.~(\ref{gr_varu})  with $u^\alpha$ and using $u_\alpha
u^\alpha=-1$ as well as Eqs.~(\ref{gr3}) and
(\ref{gr_gamma}-{\ref{gr_s}), we get
\begin{equation}
\phi_{,\alpha}u^\alpha = - \mu + Ts.
\label{gr_phi}
\end{equation}
Thus the proper time rate of change of $\phi$ along the flow is just minus the
specific Gibbs energy or chemical potential.

The importance of using Lin's $\gamma$ is clear from Eq.~(\ref{gr_varu}) . 
In the pure isentropic fluid case ($F^{\alpha\beta}=0$ and $s=$ const.), the
Khalatnikov vorticity tensor given by
\begin{equation}
\label{gr_vor2}
\omega _{\alpha \beta }=\left( \mu u_{\beta }\right) _{,\alpha }-
\left( \mu u_{\alpha }\right) _{,\beta }
=\left( \gamma \lambda _{,\beta }\right) _{,\alpha }-
\left( \gamma \lambda _{,\alpha }\right) _{,\beta }
\end{equation}
would vanish in the absence of $\gamma$, thus constraining us to discuss
only irrotational flow. This problem is well known from non
relativistic pure fluid Lagrangian theory.  Lin$^{\!\!\!\cite{lin}}$ remarked
that one can label each fluid element by its original Lagrangian
coordinate.  The requirement that this stay fixed adds an additional
constraint (``label conservation'') to the Lagrangian function, and makes
possible the description of vortical flow.   While for isentropic flow
Kelvin's theorem forbids the creation of vorticity, the flow in any given
region can be vortical due to conditions upstream.

As customary, we write $F_{\alpha\beta} = A_{\beta;\alpha} - A_{\alpha;\beta}
= A_{\beta,\alpha} -A_{\alpha,\beta}$, which ensures that the Maxwell 
Eqs.~(\ref{Maxwell2}) are automatically satisfied.  The other half,
Eqs.~(\ref{Maxwell1}), are obtained by varying with respect to the components
of the vector potential $A_\alpha$.  Because of the antisymmetry of
$F_{\alpha\beta}$, the last term of the Lagrangian, Eq.~(\ref{Lagrangian}),
can be written as $(\tau^\beta N^\alpha-\tau^\alpha N^\beta)A_{\alpha,\beta}$.
The variation of $A_\alpha$ in the corresponding term in the action
produces, after integration by parts, the term $\big[(-g)^{1/2}(\tau^\alpha
N^\beta-\tau^\beta N^\alpha)\big]_{,\beta}\, \delta A_\alpha$.  Because
for any antisymmetric tensor $t^{\alpha\beta}$,
$(-g)^{1/2}t^{\alpha\beta}{}_{;\beta} = [(-g)^{1/2}
t^{\alpha\beta}]_{,\beta}$, variation of $A_\alpha$ leads to the equation
\begin{equation}
\label{gr_mag2}
F^{\alpha \beta }{}_{;\beta } =4\pi\left( \tau^{\alpha }N^{\beta }
- \tau^{\beta }N^{\alpha }\right) _{;\beta }.
\label{tau}
\end{equation}
We see that this is just Eq.~(\ref{Maxwell1}) provided we identify the
electric current density $j^\alpha$ as
\begin{equation}
j^\alpha=\left( \tau^{\alpha }N^{\beta }
- \tau^{\beta }N^{\alpha }\right) _{;\beta }
\label{current}
\end{equation}
Since the divergence of the divergence of any antisymmetric tensor vanishes,
the charge conservation equation ($j^{\alpha}{}_{ ;\alpha} = 0$) is
satisfied automatically.  Formally Eq.~(\ref{tau}) determines the Lagrange
multiplier 4--vector $\tau^\alpha$, modulo the freedom inherent in it, as we
discuss in Sec.~III.

\subsection{MHD Euler equation in General Relativity}
\label{Euler}

We now go on to tie  the  equations of motion together to yield the MHD Euler 
equation (\ref{grmag_euler}). We begin by writing the Khalatnikov vorticity
$\omega_{\beta\alpha}$ in two forms,
\begin{equation}
\omega_{\beta\alpha} = \mu_{,\beta} u_\alpha - \mu_{,\alpha} u_\beta
 +\mu u_{\alpha;\beta} - \mu u_{\beta;\alpha},
\end{equation}
as well as by means of Eq.~(\ref{gr_varu})
\begin{eqnarray}
\label{gr_mgel2}
\omega_{\beta\alpha} &=& s_{,\beta }\eta _{,\alpha }-s_{,\alpha }\eta
_{,\beta}+\gamma _{,\beta }\lambda _{,\alpha }-\gamma _{,\alpha }\lambda
_{,\beta }
 \nonumber \\
&+&\tau^{\delta }{}_{;\beta }F_{\alpha \delta }-\tau^{\delta }{}_{;\alpha
}F_{\beta\delta }+\tau^{\delta }F_{\alpha \delta ;\beta }-\tau^{\delta
}F_{\beta
\delta;\alpha }.
\end{eqnarray}
Contracting the left hand side of the first with $ N^{\alpha } $, recalling
Eq.~(\ref{grem0}) and that by normalization $ u^{\alpha }u_{\alpha ;\beta
}=0 $ whereas $ u^{\beta }u_{\alpha ;\beta }=a_\alpha $ (recall that
$a^\alpha$ is the fluid's 4--acceleration), we get
\begin{equation}
\label{gr_mgel3}
\omega_{\beta\alpha} N^\alpha =-n\mu _{,\beta }-n\mu _{,\alpha }u^{\alpha
}u_{\beta }-n\mu a_\beta = -n h_\beta{}^\alpha\mu_{,\alpha} - n \mu a_\beta.
\end{equation}

On the other hand, contracting Eq.~(\ref{gr_mgel2}) with  $ N^{\alpha } $ and
using Eqs.~(\ref{gr_lambda}-\ref{gr_s}) and (\ref{gr3}) to drop a number of
terms, we get
\begin{equation}
\label{gr_mgel6}
\omega_{\beta\alpha} N^\alpha =-nTs_{,\beta }-\tau^{\delta }{}_{;\alpha
}F_{\beta \delta }N^{\alpha }+\tau^{\delta }F_{\alpha \delta ;\beta
}N^{\alpha}-\tau^{\delta }F_{\beta \delta ;\alpha }N^{\alpha }.
\end{equation}
By virtue of Eq.~(\ref{gr_entropy}), $-nTs_{,\beta}$ is the same as
$-n T h_\beta{}^\alpha s_{,\alpha}$.  It is convenient  to use
the thermodynamic identity $d\mu =n^{-1}\,dp+Tds$, which follows from
Eq.~(\ref{gr_mu1}) and the first law $d(\rho/n)= Tds -pd(1/n)$, to replace
$-nTs_{,\beta}$ in Eq.~(\ref{gr_mgel6})  by
$h_\beta{}^\alpha (-n\mu_{,\alpha}+p_{,\alpha})$.  Equating our two
expressions for $\omega_{\beta\alpha} N^\alpha$ gives, after a cancellation,
\begin{equation}
\label{gr_mgel9}
-\left( n\mu a_{\beta }+h_{\beta }{}^{\alpha }p_{,\alpha }\right) =
-\tau^{\delta }{}_{;\alpha }F_{\beta \delta }N^{\alpha }+\tau^{\delta
}F_{\alpha\delta ;\beta }N^{\alpha }-\tau^{\delta }F_{\beta \delta ;\alpha
}N^{\alpha }.
\end{equation}

The last two terms in this equation can be combined into a single one by
virtue of Eq.~(\ref{Maxwell2}), which can be written with
covariant as well as ordinary derivatives.  Further, by Eq.~(\ref{gr_mu1})
we may replace
$n\mu$ by $\rho+p$.  In this manner we get
\begin{equation}
\left( \rho +p\right) a_{\beta }=-h_{\beta }{}^{\alpha }p_{,\alpha }+
F_{\beta \alpha ;\delta }\tau^{\delta }N^{\alpha }+F_{\beta \delta
}\tau^{\delta }{}_{;\alpha}N^{\alpha }.
\end{equation}
The term $ \tau^{\delta }{}_{;\alpha }N^{\alpha } $ here can be replaced by
two others with help of Eq.~(\ref{gr_mag2}) if we take into
account that $N^\beta{}_{;\beta}=0$:
\begin{equation}
\label{gr_mgel12}
\left( \rho +p\right) a_{\beta }=-h_{\beta }{}^{\alpha }p_{,\alpha
}+(4\pi)^{-1} F_{\beta \delta }F^{\delta \alpha }{}_{;\alpha }+F_{\beta \alpha
;\delta }\tau^{\delta }N^{\alpha }+F_{\beta \delta }\left( \tau^{\alpha
}N^{\delta}\right) _{;\alpha }.
\end{equation}
We note that the last two terms on the right hand side combine into $\left(
F_{\beta \alpha }N^{\alpha }\tau^{\delta }\right) _{;\delta }$ which
vanishes by Eq.~(\ref{gr3}). Now substituting from the
Maxwell equations (\ref{Maxwell1}) we arrive at the final equation
\begin{equation}
\label{gr_mgel15}
\left( \rho +p\right) a_{\beta }=-h_{\beta }{}^{\alpha }p_{,\alpha }+
F_{\beta \delta }j^{\delta },
\end{equation}
which is the correct GR MHD Euler equation (\ref{grmag_euler}).  

Note that we have not used any information about $\tau^\alpha$ beyond
Eq.~(\ref{tau}); hence Euler's equation is valid for all choices of
$\tau^\alpha$, of which there are many as we shall explain.  Since we are able
to obtain all equations of motion for GR MHD from our Lagrangian density, we
may regard it as correct, and go on to look at some consequences.

\bigskip
\section{NEW CIRCULATION CONSERVATION LAW}
\label{last_sec}

\subsection{General Remarks}

Eqs.~(\ref{gr_varu}) and (\ref{gr_lambda}-\ref{gr_gamma}) lead immediately to
a law of circulation conservation for relativistic perfect isentropic MHD
flow.   Define the vector field
\begin{equation}
\label{gr_kel2}
z_\alpha \equiv {\mu}u_{\alpha }-\tau^{\beta }F_{\alpha \beta }
\end{equation}
and its associated circulation $ \Gamma $
\begin{equation}
\label{gr_kel1}
\Gamma =\oint_{{\cal C}}z_{\alpha }dx^{\alpha },
\end{equation}
where $ {\cal C} $ is a closed simply connected curve drifting with the fluid.
According to Eq.~(\ref{gr_varu}), $ z_{\alpha }=\phi _{,\alpha }+s\eta
_{,\alpha }+\gamma \lambda _{,\alpha }$. Since $ \phi _{,\alpha } $ is a
gradient, its contribution to  $\Gamma $ vanishes.  Likewise, for isentropic
flow ($ s=$ const.) the term involving $ s\eta _{,\alpha } $ makes no
contribution to $\Gamma$. Thus
\begin{equation}
\label{gr_kel3}
\Gamma =\oint_{{\cal C}}\gamma \lambda_{,\alpha} dx^\alpha = \oint_{{\cal
C}}\gamma\, d\lambda.
\end{equation}
By Eqs.~(\ref{gr_lambda}-\ref{gr_gamma}) both $\gamma$  and $ \lambda$ are
conserved with the flow.  Thus $ \Gamma  $ is a circulation which is conserved
along the flow.

In the absence of electromagnetic fields and in the nonrelativistic limit
$(\mu\rightarrow m$ where $m$ is a fluid particle's rest mass), $\Gamma $
for a curve ${\cal C}$ taken at constant time reduces to Kelvin's
circulation.  On this ground our result can be considered a
proof that a generalization of Kelvin's circulation theorem to GR MHD exists.
This conclusion goes beyond Oron's original conserved circulation in
MHD$^{\!\!\!\cite{bekenstein1}}$ in that no symmetry is necessary here for the
circulation to be conserved.

To make full use of the new conservation law to solve or simplify
problems in MHD, one must evidently know $\tau^\alpha$ explicitly.  There is
a certain amount of freedom in $\tau^\alpha$ which we have already
discussed.$^{\!\!\!\cite{bekenstein2}}$  Here it is important that the
Lagrangian density (\ref{Lagrangian}) is invariant under the addition of
$f u^\alpha$ to $\tau^\alpha$, where $f$ is an arbitrary scalar, because
$f F_{\alpha\beta} u^\alpha u^\beta$ vanishes identically.  We use this
freedom to demand that $\tau^\alpha u_\alpha=0$. We now recast the
circulation law in a form eminently suitable for use in problems with
symmetry.   

First from Eq.~(\ref{tau}) we infer that
\begin{equation}
\label{tau_sol}
\tau^{\alpha}N^{\beta} -  \tau^{\beta}N^{\alpha} = (4 \pi)^{-1} (F^{\alpha
\beta}- W^{\alpha \beta})\ ; \qquad W^{\alpha \beta}{}_{;\beta} =0
\end{equation}
where $W^{\alpha \beta}$ is an antisymmetric tensor. Since  $W^{\alpha \beta}$
must be divergenceless, it can generically be written as the dual of a curl,
\begin{equation}
\label{W}
W^{\alpha \beta} = {1 \over 2} \epsilon^{\alpha \beta \gamma \delta}
{\cal F}_{\gamma \delta}
= {}^* {\cal F}^{\alpha \beta}
\end{equation}
where ${\cal F}_{\gamma \delta} = {\cal A}_{\delta, \gamma} - {\cal
A}_{\gamma,\delta}$.   ${\cal A}_\alpha$ here is to be distinguished from the
ordinary vector potential $A_\alpha$ of $F_{\alpha\beta}$. Parenthetically we
mention that one can independently make gauge transformations of ${\cal
A}_\alpha$ and of $A_\alpha$.  This underscores the little known fact that MHD
is a theory with $U(1)\times U(1)$ gauge symmetry.

By taking the dual of  Eq.~(\ref{tau_sol}) and then contracting it with
$u^\gamma$ (remembering that twice dual is equivalent to changing
the sign) we get, with the help of Eq.~(\ref{gr2}),
\begin{equation}
B_\delta = {\cal F}_{\delta\gamma} u^\gamma.
\label{requirement}
\end {equation}
Contracting Eq.~(\ref{tau_sol}) with $u_\beta$ and
recalling that $F^{\alpha\beta}u_\beta=0$ and $\tau_\alpha u^\alpha=0$ gives
\begin{equation}
\tau^{\alpha} = (8 \pi n)^{-1} \epsilon^{\alpha\beta  \gamma
  \delta}{\cal F}_{\gamma \delta}\, u_\beta.
\label{tau_sol2}
\end{equation}

Now using Eq.~(\ref{inversion}) with $E_\alpha=0$ and Eq.~(\ref{tau_sol2})
we have
\begin{equation}
F_{\alpha\beta}\tau^\beta=(8\pi n)^{-1} \epsilon_{\alpha\beta\gamma\delta}
\epsilon^{\beta\xi\mu\nu} u^\gamma B^\delta u_\xi {\cal F}_{\mu\nu}
\label{xo}
\end{equation}
With the easily checked identity
\begin{eqnarray}
\epsilon_{\alpha\beta\gamma\delta}\epsilon^{\beta\xi\mu\nu}=
\delta ^\xi{}_\alpha \delta^\mu{}_\gamma \delta^\nu{}_\delta
-\delta ^\xi{}_\alpha \delta^\nu{}_\gamma \delta^\mu{}_\delta
+\delta ^\nu{}_\alpha \delta^\xi{}_\gamma \delta^\mu{}_\delta
\nonumber
\\
-\delta ^\mu{}_\alpha \delta^\xi{}_\gamma \delta^\nu{}_\delta
+\delta^\mu{}_\alpha \delta^\nu{}_\gamma \delta^\xi{}_\delta
-\delta ^\nu{}_\alpha \delta^\mu{}_\gamma \delta^\xi{}_\delta
\end{eqnarray}
Eq.~(\ref{xo}) reduces to
\begin{equation}
F_{\alpha\beta}\tau^\beta=(4\pi n)^{-1}\big({\cal F}_{\alpha\nu}B^\nu 
-B_\nu B^\nu u_\alpha\big)
\label{new1}
\end{equation} 
where use has been made of Eq.~(\ref{requirement}), $B_\mu u^\nu=0$ and
$u_\mu u^\mu=-1$.  The conservation law (\ref{gr_kel1}) now takes the form
\begin{equation}
\Gamma =\oint_{\cal C}\big[\chi u_\alpha - (4\pi n)^{-1} {\cal
F}_{\alpha\beta} B^\beta\big] dx^\alpha,
\label{secgam2}
\end{equation}
with $\chi\equiv \mu+(4\pi n)^{-1}B_\beta B^\beta$.  The quantity $\chi$ plays
an important role in  Oron's generalization of Kelvin's
theorem to MHD.$^{\!\!\!\cite{bekenstein1}}$  

\subsection{Example: MHD Flow with Three Symmetries}

Since ${\cal F}_{\alpha\beta}$ is not known explicitly, we cannot work out
the conserved circulation without further work.  Here we shall
make some progress in this direction in the case of flow which is both
stationary, and  possesses two additional spatial symmetries (these last we
assume {\it not\/} to be of the angular type).  This means the physical
quantities such as $u^\alpha$, $N^\alpha$, $j^\alpha$ or $B^\alpha$ are
unchanged upon being Lie dragged along either of the three Killing vectors.
We cannot automatically require the same of ${\cal F}_{\alpha\beta}$ because
it is not a directly measurable quantity.  However, as mentioned in our
earlier work,$^{\!\!\!\cite{bekenstein2}}$ we can make a transformation $
{\cal F}_ {\alpha \beta}\rightarrow {\cal F}_ {\alpha \beta} + f_ {\alpha
\beta}$, where $f_{\alpha\beta}$ is a curl and orthogonal to $u^\beta$,
without changing the values of $j^\alpha$ or $B^\alpha$; this transformation 
at most adds to $\Gamma$ a conserved quantity leaving it conserved.  We shall
assume here that by means of such a transformation we can make ${\cal
F}_{\alpha\beta}$ share the symmetries of $j^\alpha$ or $B^\alpha$.

The following remarks apply when we choose the coordinates $x^2$ and $x^3$ to
extend along the integral curves of the two spatial Killing vectors; by
assumption these curves are noncompact, and so are $x^2$ and $x^3$.  We also
assume $x^1$ is noncompact. The most general form for the ``vector
potential'' ${\cal A}_\alpha$ for which ${\cal F}_{\alpha\beta}$ is
independent of  $x^0$, $x^2$ and $x^3$, is (here we sacrifice manifest
covariance in order to make the symmetries manifest)
\begin{equation}
{\cal A}_\alpha =  x^0\hat\Phi_{,\alpha} + x^2\hat\Psi_{,\alpha} +
x^3\hat\Xi_{,\alpha} +{\cal V}_\alpha
\label{4-potential}
\end{equation}      
where $\hat\Phi, \hat\Psi$ and $\hat\Xi$ are each a linear combination of
$x^0, x^2$ and $x^3$ with constant coefficients plus a function of the
nontrivial coordinate $x^1$ only, while the components of the ``vector''
${\cal V}_\alpha$ also depend only on $x^1$. Thus
\begin{eqnarray}
{\cal F}_{01}&=&(\hat\Phi-{\cal V}_0)_{,1}\equiv \Phi_{,1}
\label{F01}
\label{first}
\\
{\cal F}_{21}&=&(\hat\Psi-{\cal V}_2)_{,1}\equiv \Psi_{,1}
\label{second}
\\
{\cal F}_{31}&=&(\hat\Xi-{\cal V}_3)_{,1}\equiv \Xi_{,1}
\label{third}
\end{eqnarray}
while ${\cal F}_{02}, {\cal F}_{03}$ and ${\cal F}_{23}$ are all strictly
constant, and thus stand for global parameters of the flow.

By means of Eq.~(\ref{requirement}) we may now compute the components
of $B_\alpha$:
\begin{eqnarray}
B_0&=\Phi_{,1}u^1+{\cal F}_{02}u^2+{\cal F}_{03}u^3
\\
B_1&=-\Phi_{,1}u^0-\Psi_{,2}u^2-\Xi_{,1}u^3
\\
B_2&=\Psi_{,1}u^1+{\cal F}_{20}u^0+{\cal F}_{23}u^3
\\
B_3&=\Xi_{,1}u^1+{\cal F}_{30}u^0+{\cal F}_{32}u^2
\end{eqnarray}
Solving these last for the derivatives
of $\Phi, \Psi$ and $\Xi$ we get from (\ref{first})-(\ref{third})    
\begin{eqnarray}
{\cal F}_{01}&=&(B_0-{\cal F}_{02}u^2-{\cal F}_{03}u^3)/u^1
\label{1}
\\
{\cal F}_{21}&=&(B_2+{\cal F}_{02}u^0-{\cal F}_{23}u^3)/u^1
\label{2}
\\
{\cal F}_{31}&=&(B_3+{\cal F}_{03}u^0+{\cal F}_{23}u^2)/u^1
\label{3}
\end{eqnarray}

We now calculate the quantity appearing last in Eq.~(\ref{secgam2}) with the
help of Eqs.~(\ref{1})-(\ref{3}).  If temporarily we take ${\cal
F}_{02}={\cal F}_{03}=0$, we get
\begin{eqnarray}
{\cal F}_{\alpha\beta} B^\beta dx^\alpha = (B^1/u^1)(B_0 dx^0
+B_2 dx^2+B_3 dx^3) - (B_0B^0 + B_2 B^2 + B_3 B^3) (dx^1/u^1)
\nonumber
\\
 +({\cal F}_{23}/u^1)\big[(u^3 B^2 - u^2 B^3)dx^1+(u^1 B^3 - u^3 B^1)dx^2+(u^2
B^1 - u^1 B^2)dx^3\big]
\end{eqnarray}
This is much simplified by adding $B_1 B^1 (dx^1/u^1)$ to the first term and
subtracting it from the second term.  In addition, one can unify the
terms in square brackets by employing the Levi-Civita tensor.  Putting
all this together and using the expression (\ref{inversion}) we have
\begin{equation}
\Gamma=\oint_{\cal C}\left[\chi u_\alpha + (4\pi n u^1)^{-1}\left( B_\beta
B^\beta\delta^1{}_\alpha -B^1 B_\alpha  + ({\cal
F}_{23}/\sqrt{-g})F_{0\alpha}\right) \right]dx^\alpha
\label{lastgamma}
\end{equation}

We now show that the last term in the integrand does not contribute.  First
of all we can take the constant ${\cal F}_{23}$ out of the integral.  Next
we realize that the law of particle number conservation (\ref{gr4}) and the
assumed symmetries tell us that $-C^{-1}\equiv \sqrt{-g}nu^1$ does not depend
on $x^1$ either, and so can also be taken out of the integral.  We are left
with a term proportional to $C\oint F_{0\alpha}dx^\alpha$.  Now, just like
${\cal F}_{\alpha\beta}$, $F_{\alpha\beta}$ derives from a vector potential,
and must have all the symmetries we have assumed.  We can thus take its vector
potential $A_\alpha$ to have the form (\ref{4-potential}), but with new
functions $\Phi', \Psi', \Xi'$ ({\it sans\/} circumflex), each of which is,
again, a linear combination of $x^0, x^2$ and $x^3$ with constant
coefficients plus a function of the nontrivial coordinate $x^1$ only.  It
follows that $F_{0\alpha}dx^\alpha=\Phi'_{,1}dx^1 + {\rm const.} dx^2+ {\rm
const.} dx^3$.  But this is a perfect differential; hence the term
proportional to
${\cal F}_{23}$ in Eq.~(\ref{lastgamma}) vanishes.
If we now reinstate ${\cal F}_{02}$ and ${\cal F}_{03}$ we find that they
contribute to $\Gamma$ terms proportional to $C{\cal F}_{02}\oint
F_{3\alpha}dx^\alpha$ and $C{\cal F}_{03}\oint
F_{2\alpha}dx^\alpha$, both of which are found to vanish by reasoning
analogous to the above.

By the symmetries the term $B_\beta B^\beta(nu^1)^{-1}$ in the integrand of
Eq.~(\ref{lastgamma}) can only depend on $x^1$.  It is integrated over $x^1$
only, once forward and once backward because ${\cal C}$ is a closed curve. 
Hence this term makes no contribution to $\Gamma$.  Further, it is a
consequence of Euler's equation (\ref{grmag_euler}) that $(\mu
B^\alpha)_{;\alpha}=0$.$^{\!\!\!\cite{bekenstein1}}$   Only the radial
derivative survives here, so that we have $\sqrt{-g}B^1=D/\mu$ with $D$ a
constant, a further global parameter of the flow.  We thus obtain the final
form of the conserved circulation:
\begin{equation}
\Gamma=\oint_{\cal C}\left[\chi u_\alpha + DC(4\pi \mu)^{-1}B_\alpha\right]
dx^\alpha
\end{equation}
This is exactly Oron's original conserved
circulation$^{\!\!\!\cite{bekenstein1}}$ for two symmetries (we have here
defined $C$ and $D$ to correspond to the two quantities with the same names
which are conserved along streamlines for the case of two
symmetries$^{\!\!\!\cite{bekenstein1}}$).  Thus the imposition of a third
symmetry does not cause any changes in $\Gamma$.  In future work we shall
endeavor to recover Oron's conserved quantity in the presence of two
symmetries from the present approach, as well as to explore the conserved
circulation in stationary flow with no spatial symmetry.  The general case of
no symmetries is a more distant goal.

\bigskip
\section*{ACNOWLEDGMENTS}This work is supported by a grant from the Israel
Science Foundation, which was established by the Israel National Academy of
Sciences.      

\end{document}